\documentclass[12pt,aps,pra,groupedaddress,amssymb,amsfonts,nofootinbib,tightenlines,showpacs]{revtex4} 
\usepackage{graphicx} 

\graphicspath{{./}{graphics/}}
\usepackage{times,color}
\definecolor{purple}{rgb}{0.625,0.125,0.9375}
\newcommand{\ignore}[1]{}

\definecolor{grey}{rgb}{0.8,0.8,0.8}

\ignore{
rm -R /tmp/onns; mkdir /tmp/onns
cp `texfls onns.log | perl -e '$a=<>; $a =~ s:/\S*(revtex4|natbib)\S*(\s|$)::g; print "$a\n"'` /tmp/onns
find /tmp/onns -type f -name '*.pdf' -exec pdftops -eps {} \; -exec rm {} \;
find /tmp/onns -type f -name '*.jpg' -exec perl -e '$f=shift; $g=f$; $g=~s/.jpg/.eps/; exec("convert $f $g");' {} \; -exec rm {} \;
perl -i -n -e 's/^

(cd /tmp/onns; tar czvf onns.tar.gz *)
}

\ignore{
rm -R /tmp/onns; mkdir /tmp/onns
cp `texfls onns.log | perl -e '$a=<>; $a =~ s:/\S*(revtex4|natbib)\S*(\s|$)::g; print "$a\n"'` /tmp/onns
find /tmp/onns -type f -name '*.pdf' -exec pdftops -eps {} \; -exec rm {} \;
find /tmp/onns -type f -name '*.jpg' -exec perl -e '$f=shift; $g=f$; $g=~s/.jpg/.eps/; exec("convert $f $g");' {} \; -exec rm {} \;
perl -i -n -e 's/^

(cd /tmp/onns; tar czvf onns.tar.gz *)
}

\newcommand{\sysfnt}{\mathsf}

\newcommand{\ket}[1]{{|}{#1}{\rangle}}
\newcommand{\bra}[1]{{\langle}{#1}{|}}
\newcommand{\braket}[2]{{\langle}{#1}{|}{#2}{\rangle}}
\newcommand{\ketbra}[2]{\ket{#1}\bra{#2}}
\newcommand{\kets}[2]{{|}{#1}{\rangle}_{{}_{\!\!\scriptstyle{\sysfnt{#2}}}}}
\newcommand{\bras}[2]{{}^{\scriptstyle\sysfnt{ #2}}\!{\langle}{#1}{|}}

\newcommand{\ketbras}[3]{\kets{#1}{#3}\!\!\bras{#2}{#3}}
\newcommand{\slb}[2]{{{#1}^{({\sysfnt{#2}})}}}

\newcommand{\cmplx}{\mathbb{C}}

\newcommand{\cA}{{\cal A}}

\newcommand{\cD}{{\cal D}}
\newcommand{\cE}{{\cal E}}

\newcommand{\cG}{{\cal G}}
\newcommand{\cH}{{\cal H}}
\newcommand{\cI}{{\cal I}}
\newcommand{\cJ}{{\cal J}}
\newcommand{\cN}{{\cal N}}
\newcommand{\cO}{{\cal O}}

\newcommand{\cR}{{\cal R}}
\newcommand{\cS}{{\cal S}}

\newcommand{\cV}{{\cal V}}
\newcommand{\cZ}{{\cal Z}}

\renewcommand{\tensor}{\otimes}
\newcommand{\trace}{\mbox{tr}}

\newcommand{\one}{\mathbf{I}}
\newcommand{\zero}{\mathbf{0}}

\newcommand{\suchthat}{\mbox{$\mathbf{:}$}}
\newcommand{\Null}{\mbox{Null}}
\newcommand{\Supp}{\mbox{Supp}}

\newtheorem{Theorem}{Theorem}
\newtheorem{Lemma}[Theorem]{Lemma}
\newtheorem{Corollary}[Theorem]{Corollary}

\newcommand{\proof}{\paragraph*{Proof.}}
\newcommand{\proofof}[1]{\paragraph*{Proof of Thm.#1.}}
\newcommand{\proofofx}[1]{\paragraph*{Proof of #1.}}
\newcommand{\qed}{\hspace*{\fill}\rule{2.5mm}{2.5mm}%
\vspace*{8pt}\par}

\begin{document}
\title{On Protected Realizations of Quantum Information}
\author{E. Knill}
\email[]{knill@boulder.nist.gov}
\affiliation{National Institute of Standards and Technology} 

\date{\today}
\begin{abstract}
There are two complementary approaches to realizing quantum
information so that it is protected from a given set of error
operators.  Both involve encoding information by means of subsystems.
One is initialization-based error protection, which involves a quantum
operation that is applied before error events
occur~\cite{knill:qc1999b}.  The other is operator quantum error
correction, which uses a recovery operation applied after the
errors~\cite{kribs:qc2004a}. Together, the two approaches make it
clear how quantum information can be stored at all stages of a process
involving alternating error and quantum operations. In particular,
there is always a subsystem that faithfully represents the desired
quantum information.  We give a definition of faithful realization of
quantum information and show that it always involves subsystems. This
justifies the ``subsystems principle'' for realizing quantum
information~\cite{viola:qc2000c}.  In the presence of errors, one can
make use of noiseless, (initialization) protectable, or
error-correcting subsystems.  We give an explicit algorithm for
finding optimal noiseless subsystems by refining the strategy given
in~\cite{choi:qc2005b}. Finding optimal protectable or
error-correcting subsystems is in general difficult. Verifying that a
subsystem is error-correcting involves only linear
algebra~\cite{kribs:qc2004a,kribs:qc2005a,nielsen:qc2005b}. We discuss
the verification problem for protectable subsystems and reduce it to a
simpler version of the problem of finding error-detecting codes.
\end{abstract}

\ignore{ 

There are two complementary approaches to realizing quantum
information so that it is protected from a given set of error
operators.  Both involve encoding information by means of subsystems.
One is initialization-based error protection, which involves a quantum
operation that is applied before error events occur.  The other is
operator quantum error correction, which uses a recovery operation
applied after the errors. Together, the two approaches make it clear
how quantum information can be stored at all stages of a process
involving alternating error and quantum operations. In particular,
there is always a subsystem that faithfully represents the desired
quantum information.  We give a definition of faithful realization of
quantum information and show that it always involves subsystems. This
justifies the "subsystems principle" for realizing quantum
information. In the presence of errors, one can make use of noiseless,
(initialization) protectable, or error-correcting subsystems.  We give
an explicit algorithm for finding optimal noiseless
subsystems. Finding optimal protectable or error-correcting subsystems
is in general difficult. Verifying that a subsystem is
error-correcting involves only linear algebra. We discuss the
verification problem for protectable subsystems and reduce it to a
simpler version of the problem of finding error-detecting codes.

}


\maketitle

\section{Introduction}
\label{sect:introduction}

According to quantum information theory, quantum information is
represented by states of a number of qudits, which are idealized
$d$-level quantum systems. Here, we consider finite quantum
information and do not explicitly refer to the underlying
tensor-product structure of the state space of more than one
qudit. Thus, we consider quantum information represented by the states
of an ideal system $\sysfnt{I}$ whose state space is determined by a
Hilbert space $\cH_{\sysfnt{I}}$ of finite dimension $N$.  In order to
realize quantum information in a physical system $\sysfnt{P}$, two
problems need to be solved. The first is to determine the ways in
which the state space of $\sysfnt{P}$ can usefully encode the desired
quantum information, the second is to determine which among these ways
can be best protected from decoherence due to the dynamics of
$\sysfnt{P}$ and its interactions. If quantum information is intended
for use in quantum algorithms, a third problem is to ensure that the
dynamics of the system and its interactions can be used to implement
the desired quantum gates. Here we consider the first two
problems. The mathematical conventions for objects and notations are
explained at the end of the Introduction.

\subsection{Encoding Quantum Information and the Subsystems Principle}
\label{sect:encoding}

A general method for realizing or encoding quantum information is as a
subsystem of $\cH_{\sysfnt{P}}$~\cite{viola:qc2000c}. This method
involves a decomposition of $\cH_{\sysfnt{P}}$ as
\begin{equation}
\cH_{\sysfnt{P}} = \left(\cH_{\sysfnt{I'}}\tensor\cH_{\sysfnt{S}}\right)
                     \oplus
                     \cH_{\sysfnt{R}},
\label{eq:subenc}
\end{equation}
where $\oplus$ denotes the orthogonal sum of Hilbert spaces.  This
decomposition identifies $\sysfnt{I'}$ as a subsystem of $\sysfnt{P}$,
written as $\sysfnt{I'}\hookrightarrow\sysfnt{P}$.  We use ``primes''
(as in $\sysfnt{I'}$) for systems with identical state spaces. With
this convention, Eq.~(\ref{eq:subenc}) yields a representation of the
states of the ideal quantum information system $\sysfnt{I}$ in a
subsystem of $\cH_{\sysfnt{P}}$.  We say that $\sysfnt{I}$ is
``encoded'' in $\sysfnt{P}$ and call the decomposition of
Eq.~(\ref{eq:subenc}) a ``subsystem encoding''. For lack of a better word,
we refer to $\sysfnt{S}$ as the ``cosubsystem'' of $\sysfnt{I'}$ in
$\sysfnt{P}$.  $\sysfnt{R}$ is the ``remainder system''.  An example
of such an encoding is that of a vibrational qubit of a single ion
trapped in a one-dimensional harmonic potential. In this case the state
space is spanned by the internal and vibrational levels of the
ion. The space $\cH_{\sysfnt{I'}}\tensor \cH_{\sysfnt{S}}$ is formed
from the first two vibrational levels and the internal levels of the
ion, respectively. The space $\cH_{\sysfnt{R}}$ consists of states
with more than two vibrational quanta irrespective of the internal
state of the ion. The familiar cases of such encodings have the
property that $\cH_{\sysfnt{I'}}$ belongs to a physically meaningful
degree of freedom. However, for the purpose of protecting against
errors, it is usually necessary to use ``entangled'' encodings. An
example is the noiseless qubit encoded in three spin-$1/2$ particles
subject to collective decoherence~\cite{knill:qc1999b}.  A feature of
subsystem encoding is that the states of $\sysfnt{I}$ are not uniquely
encoded as states of $\sysfnt{P}$. This is because any change of state
of the cosubsystem $\sysfnt{S}$ does not affect the states of
$\sysfnt{I'}$.

Subsystem encodings of $\sysfnt{I}$ in $\sysfnt{P}$ are equivalent to
$\dagger$-preserving isomorphic embeddings of $B(\cH_{\sysfnt{I}})$
into $B(\cH_{\sysfnt{P}})$~\cite{knill:qc1999b,zanardi:qc1999c}.  Here
$B(\cH)$ denotes the set of (bounded) operators on $\cH$.  In
particular, given the subsystem encoding of Eq.~(\ref{eq:subenc}),
$B(\cH_{\sysfnt{I}})$ is isomorphic to the algebra of operators of the
form $A\tensor \one \oplus \zero$ with $A$ acting on
$\cH_{\sysfnt{I'}}$, $\one$ on $\cH_{\sysfnt{S}}$ and $\zero$ on
$\cH_{\sysfnt{R}}$, where the operators are transported to
$\cH_{\sysfnt{P}}$ via the isomorphism implicit in the subsystem
decomposition as needed. Conversely, if $\cA$ is a subalgebra of
bounded operators on $\cH_{\sysfnt{P}}$ and $\cA$ is
$\dagger$-isomorphic to $B(\cH_\sysfnt{I})$, then there is a unique
subsystem encoding such that the operators of $\cA$ are the operators
of the form $A\tensor\one \oplus\zero$ as above.

Are there ways of encoding quantum information that do not involve a
subsystem encoding?  In an attempt to answer this question, it is
worth considering other prescriptions for encoding quantum
information. There are two operationally defined ways of
characterizing encoded information. The first is by a traditional
encoding operation that isometrically embeds $\cH_{\sysfnt{I}}$ into
$\cH_{\sysfnt{P}}$. (Isometries are linear maps preserving the inner
product.)  This is the prescription used in the traditional theory of
quantum error correction and corresponds to a subsystem encoding with
trivial cosubsystem.  In this case, the subspace
$\cH_{\sysfnt{I'}}\tensor\cH_{\sysfnt{S}} = \cH_{\sysfnt{I'}}$ of
$\cH_{\sysfnt{P}}$ is known as a ``quantum code''. That this is
inadequate is apparent when one considers enlarging $\sysfnt{P}$ by
other relevant degrees of freedom. Subspaces also fail to capture the
location of quantum information in realistic error-control settings,
in particular fault-tolerant quantum computing.  This is because in
practice, error control never results in restoration of the encoded
quantum information to any fixed quantum code. Assuming that this is a
requirement leads to the conclusion that fault tolerance is not
possible~\cite{alicki:qc2004b}.

The second operational realization of quantum information involves
specifying an ideal decoding procedure. Such a decoding procedure
adjoins $\sysfnt{I}$ and (possibly) an ancilla system $\sysfnt{A}$ to
$\sysfnt{P}$, where $\sysfnt{I}$ and $\sysfnt{A}$ are in specified
initial states $\kets{0}{I}$ and $\kets{0}{A}$. The total state space
is determined by
$\cH_{\sysfnt{P}}\tensor\cH_{\sysfnt{I}}\tensor\cH_{\sysfnt{A}}$.  The
decoding operation is a unitary operator on this state space. After it
is applied, the desired quantum information resides in $\sysfnt{I}$.
The decoding-based view of quantum information has been used
successfully in analyses of fault-tolerant quantum architectures (see,
for example,~\cite{aliferis:qc2005b}). To connect the decoding-based
realization of quantum information to subsystems, note that the
decoding operation is, in effect, an isometry from an extended space
$\cH_{\sysfnt{P_e}}\oplus\cH_{\sysfnt{T}}$ to
$\cH_{\sysfnt{I}}\tensor\cH_{\sysfnt{U}}$. Here we have identified
$\cH_{\sysfnt{P_e}}$ with
$\cH_{\sysfnt{P}}\tensor\kets{0}{I}\tensor\kets{0}{A}$ and
$\cH_{\sysfnt{U}}$ with $\cH_{\sysfnt{A}}\tensor\cH_{\sysfnt{P}}$.
Decoding-based realization is therefore equivalent to subsystem
encoding in an extension of the physical state space, where the
extension need not be physically meaningful. One can consider
generalizing decoding-based realizations by means of isometries
that provide the identification
\begin{equation}
\cH_{\sysfnt{P}}\tensor\cH_{\sysfnt{I}}\oplus\cH_{\sysfnt{T}}
=\cH_{\sysfnt{I'}}\tensor\cH_{\sysfnt{S}}\oplus\cH_{\sysfnt{R}},
\end{equation}
which is obtained if the decoding operation also involves additional
physical systems in unspecified initial states.  Although this is more
general than subsystem encoding, most such isomorphisms do not result
in quantum information that can be considered to be faithfully encoded
in system $\sysfnt{P}$.  We resolve this problem at the end of
Section~\ref{sect:faithful} by pairing decoding and encoding
operations.

A third approach to encoding of quantum information uses operators
to characterize quantum information and makes it possible to give a
reasonable definition of ``faithful encoding of $\sysfnt{I}$ in
$\sysfnt{P}$''. We give such a definition at the beginning of
Sect.~\ref{sect:faithful} and prove that every such encoding is
associated with a subsystem. The intuition is that the states of
$\sysfnt{I}$ are characterized by the expectations of a linearly
closed set of observables $\cO$.  To ensure the correct dynamics of
these states, the complex multiples of the observables should form an
algebra $\dagger$-isomorphic to $B(\cH)$.  Thus, one expects that
faithful encoding of quantum information requires identifying a
$\dagger$-closed subalgebra $\cA$ of $B(\cH_{\sysfnt{P}})$ that is
isomorphic to $B(\cH_{\sysfnt{I}})$. If this has been done, then the
representation theory of $\dagger$-closed algebras uniquely identifies
a decomposition $\cH_{\sysfnt{P}}=
\cH_{\sysfnt{I'}}\tensor\cH_{\sysfnt{S}}\oplus\cH_{\sysfnt{R}}$ such
that $\cA$ consists of all the operators acting only on
$\cH_{\sysfnt{I'}}$. The results of Sect.~\ref{sect:faithful} provide
support for the ``subsystems principle'' for realizing quantum
information:

\noindent\textbf{The subsystems principle:}
\emph{
Any faithful representation of quantum information in
a physical system requires that at every point in time
there are identifiable subsystems encoding the desired
quantum information.
}

The subsystem principle is powerful, but it is worth noting that it is
sometimes convenient to use realizations of quantum information that
do not satisfy this principle perfectly.  For example, in optical
quantum computing with ``cat'' states, it is convenient to represent
the logical states of qubits by non-orthogonal coherent
states~\cite{ralph:qc2001b,ralph:qc2003a}. Another example is the
study of ``initialization-free'' decoherence-free subsystems, where
the probability amplitude of the encoded information may be less than
$1$ and the nature of the remaining amplitude must be taken into
account~\cite{shabani:qc2005a}. 

\subsection{Protecting Quantum Information}
\label{sect:protect}

In most physical settings there are sources of errors that can affect
encoded quantum information. Ideally, we would like exact knowledge of
the error behavior of a physical system under all circumstances in
which it is used. Since this knowledge is usually unavailable, one of
a number of idealizing assumptions can be made. In the context of
quantum channels, or when unwanted interactions are expected to have
weak temporal correlations, we assume errors to be due to a known
markovian process (in the continuous time setting) or a known quantum
operation (in the discrete time setting).  Both may be described by
a collection of possible error events $\cE=\{E_i\}$.  In general,
the goal of quantum error control is to find quantum information
subsystems for which the effects of the markovian process or quantum
operation can be suppressed to the largest extent possible.  Because
the exact nature of the errors is usually not known, this goal is
typically difficult to pursue. To make the task more tractable, we can
consider only those errors that are expected to be likely and look for
subsystems that allow for ``good'' protection against such errors. We
can then bound the effect of other errors by making estimates of their
maximum probability (or amplitude) of occurrence.

In this paper, we focus on subsystems that enable perfect protection
against a fixed set of errors $\{E_i\}_i$, with or without active
intervention. Because of the linearity of quantum mechanics, perfect
protection against the $E_i$ implies perfect protection against any
error in the linear span $\cE$ of the $E_i$.  Subsystems whose states
are unaffected by the errors are known as ``noiseless'' or
``decoherence free'' subsystems~\cite{onns1} and were introduced
in~\cite{knill:qc1999b,zanardi:qc1999c} in the context of
$\dagger$-closed $\cE$ (or the $\dagger$-closure of a
non-$\dagger$-closed $\cE$), in which case they can be characterized
by irreducible representations of the commutant of $\cE$,
which is the set of operators that commute with all members of
$\cE$. In general, noiseless subsystems are not as easily
characterized. In~\cite{kempe:qc2001a} an explicit characterization of
noiseless subsystems for any $\cE$ is obtained.  This characterization
is readily seen to be equivalent to the statement that the subsystem
$\sysfnt{I'}$ of the decomposition $\cH_{\sysfnt{P}} =
\cH_{\sysfnt{I'}}\tensor\cH_{\sysfnt{S}}\oplus\cH_{\sysfnt{R}}$ is
noiseless if and only if the restriction of the $E_i$ to the subspace
$\cH_{\sysfnt{I'}}\tensor\cH_{\sysfnt{S}}$ acts as
$\slb{\one}{I'}\tensor \slb{E'_i}{S}$.  Several equivalent
characterizations for when $\cE$ is the span of the operation elements
of a specific quantum operation were obtained
in~\cite{kribs:qc2004a,kribs:qc2005a}.  These characterizations do not
directly address the question of how one can computationally search
for noiseless subsystems.  A strategy for this search was offered
in~\cite{choi:qc2005b}.  This strategy requires finding
$\cE$-invariant subspaces and decomposing them into the canonical
subsystems associated with the irreducible representations of a fixed-point algebra for a
quantum operation whose operation elements span $\cE$.  In
Sect.~\ref{sect:ns} we develop this strategy into an algorithm that
does not require explicit constructions of algebras other than that
generated by $\cE$. The mathematical structure of algebras over the
complex numbers plays a crucial role. Interestingly, if there exists a
quantum operation whose operational elements span $\cE$, then the
algorithm simplifies substantially and is efficient in the dimension
of the Hilbert space. Note that there is no a priori requirement that
the likely errors included in $\cE$ be derived from a quantum
operation.  However, in most cases $\cE$ does satisfy this condition.
To ensure that this condition holds, one can add $\one-\lambda\sum_i
E_i^\dagger E_i$ for a sufficiently small $\lambda$, although the
choice of spanning set $E_i$ and $\lambda$ may affect the availability
of large-dimensional noiseless subsystems.

When no noiseless subsystem of sufficiently large dimension can be
found, it is necessary to use active intervention to protect encoded
quantum information. The idealized setting for active intervention
involves alternating steps consisting of error events $E_i$ and a
quantum operation $\cR$ that ensures that the errors do not affect the
encoded information.  An operation $\cR$ with this property is known
either as a ``recovery'' or as an ``initialization'' operation,
depending on context.  According to the subsystems principle, there
must be two subsystems, one in which quantum information resides after
error events but before $\cR$ is applied, and another after $\cR$ is
applied. We call the first a ``protectable'' subsystem. The second is
known as an ``error-correcting'' subsystem. Provided the encoded
quantum information has been successfully protected, both subsystems
are noiseless.  The first is noiseless for the products $E_iR_j$,
where the $R_j$ are the operation elements of $\cR$, whereas the
second is noiseless for the operators $R_jE_i$.  Protectable
subsystems are defined (but not named) in~\cite{knill:qc1999b}, where
it was shown how to determine the protectable subsystem in the case
where the error-correcting subsystem is a quantum code, that is, the
cosubsystem is one-dimensional. Error-correcting subsystems are the
main feature of operator quantum error
correction~\cite{kribs:qc2004a,kribs:qc2005a} and directly generalize
traditional error-correcting codes.

Knowledge of the protectable subsystem and the error-correcting
subsystem associated with a recovery/initialization operation and the
relationship between the two helps us to understand how quantum
information is stored at all times. An advantage of the protectable
subsystem is that in many cases it is a simple extension of the
error-correcting subsystem. That is, the former's cosubsystem is a
consistent extension of the latter's cosubsystem. As a result, the
observables associated with the protectable subsystem induce the
correct observables on the error-correcting subsystem.  This implies
that for the purpose of identifying the current value of the stored
quantum information, it suffices to know the protectable subsystem,
regardless of whether the last event was a recovery operation or an
error. Examples of this situation are stabilizer codes with decoding
algorithms based on syndrome extraction. It is readily verified that
the associated protectable subsystem contains the stabilizer code as a
subspace where the cosubsystem is in a particular state. In
particular, this property holds for the stabilizer-based
error-correcting subsystems identified in~\cite{poulin:qc2005a} and
used to simplify Shor's 9-qubit one-error-correcting
code~\cite{bacon:qc2005b}, except that the error-correcting subsystem
is defined by a subspace of the protectable subsystem's
cosubsystem. In general, this relationship between protectable and
error-correcting subsystems always holds if $\one\in\cE$. It becomes
particularly useful in the context of fault-tolerant quantum
computation, where the recovery operation and error events can no
longer be easily separated. In this case the ideal error-correcting
codes or subsystems associated with a scheme are typically not where
quantum information resides. It resides in the protectable subsystems
of the scheme. Note that in this setting it is usually the case that
the subsystems containing quantum information vary in time. This
happens, for example, when teleportation is used for error correction,
when quantum information is stored in memory versus being actively
manipulated, and in cluster-state-based schemes as part of the
model~\cite{raussendorf:qc2001a}.

An advantage of error-correcting subsystems is that there are simple
criteria and algorithms for determining whether there exists an
associated recovery operation for which it becomes
noiseless~\cite{kribs:qc2004a,kribs:qc2005a,nielsen:qc2005b}.  Not
having to specify the recovery operation simplifies the search for
subsystems suitable for protecting quantum information and makes it
natural to talk about error-correcting subsystems without specifying
the recovery operation. The same cannot be said for protectable
subsystems. In Sect.~\ref{sect:initialization} we partially remedy
this situation by reducing the problem of determining whether a
subsystem is protectable to a number of other problems not involving
the existence of a quantum operation.

\subsection{Conventions}
Capital letters in sans-serif font such as $\sysfnt{A,...,H,...,P}$
are used to label quantum systems.  The state space of a system
$\sysfnt{A}$ is determined by a Hilbert space, denoted by
$\cH_{\sysfnt{A}}$. We label states according to the quantum system
they belong to. For example, $\kets{\psi}{A}$ is a pure state of
$\sysfnt{A}$ and $\slb{\rho}{A}$ is a density matrix for
$\sysfnt{A}$. The tensor product symbol $\tensor$ may be omitted in
tensor products of labeled states and operators.  We frequently
consider instances of identical state spaces realized by and in
different systems. We use primes to distinguish the different systems
with identical state spaces. Thus, $\sysfnt{I}$, $\sysfnt{I'}$ and
$\sysfnt{I''}$ are systems whose state spaces are identified via
implicit isometries, which are inner-product-reserving linear maps.
In particular, a state $\kets{\psi}{I}$ of $\sysfnt{I}$ is identified
with the states $\kets{\psi}{I'}$ and $\kets{\psi}{I''}$ of systems
$\sysfnt{I'}$ and $\sysfnt{I''}$. One way to interpret this is to
consider $\psi$ as a symbol labeling a vector in an appropriate
Hilbert space $\cH$ and $\ket{\psi}\mapsto\kets{\psi}{I}$ as the
isometry identifying $\cH$ and $\cH_{\sysfnt{I}}$.  We use the
equality symbol ``$=$'' not just to denote strict mathematical
equality but also for identifying objects which are equal via an
isomorphism. The isomorphisms involved are defined only implicitly,
provided the meaning is clear. For a Hilbert space $\cH$, $B(\cH)$
denotes the algebra of operators of $\cH$.  $U(\cH)$ denotes the group
of unitary operators of $\cH$. In this work, all state spaces are
finite dimensional.

\section{Faithful Encodings of Quantum Information}
\label{sect:faithful}

To formalize the idea of ``faithful encoding'' we consider more
general ways of encoding quantum information.  A faithful encoding of
$\sysfnt{I}$ in $\sysfnt{P}$ is a map $D$ from density operators $\rho$ on
$\cH_{\sysfnt{I}}$ to non-empty sets of density operators on
$\cH_{\sysfnt{P}}$ together with a map $O$ from observables (hermitian
operators) $A$ of $\cH_{\sysfnt{I}}$ to non-empty sets of observables of
an extension $\cH_{\sysfnt{Q}}$ of $\cH_{\sysfnt{P}}$ that satisfies three
faithfulness requirements:
\begin{itemize}
\item[1.] Statics: For all $\sigma\in D(\rho)$ and $X\in O(A)$,
\begin{equation}
\label{eq:statics}
\trace(\sigma X)=\trace(\rho A).
\end{equation}
This requirement ensures that we can identify the expectation
values of faithfully encoded states.
\item[2.] Unitary dynamics:
For all $\sigma\in D(\rho)$ and $X\in O(A)$,
\begin{equation}
\label{eq:udynamics}
e^{-iX}\sigma e^{iX} \in D(e^{-iA}\rho e^{iA}).
\end{equation}
With this requirement satisfied, we can evolve the states using
conventional quantum control so that the evolved states are consistent
with the first requirement.
\end{itemize}
For the next requirement, extend the domain of $D$ to all positive
semidefinite operators by defining $D(\zero) = \{\zero\}$ and for
$\rho\not=\zero$, $D(\rho) = \trace(\rho)D(\rho/\trace(\rho))$.  For
an operator $Z$, let $\Pi(Z,\lambda)$ be the projector onto the
$\lambda$-eigenspace of $Z$, or, equivalently, the projector onto the
null space of $Z-\lambda$. For $\lambda$ not in the spectrum of $Z$,
the projector is $\zero$.
\begin{itemize}
\item[3.] Measurement dynamics: For all $\sigma\in D(\rho)$ and $X\in
O(A)$ and $\lambda$ real, 
\begin{equation}
\label{eq:mdynamics}
\Pi(X,\lambda)\sigma\Pi(X,\lambda)\in
D(\Pi(A,\lambda)\rho\Pi(A,\lambda)). 
\end{equation}
Faithful measurement dynamics ensures that projective
measurements can be implemented correctly.
\end{itemize}

The support of a positive semidefinite Hermitian operator $\rho$ is
the span of its non-zero eigenvalue eigenvectors and is denoted by
$\Supp(\rho)$. For a set of such operators $D$, $\Supp(D)$ is the
span of the supports of the members of $D$.

\begin{Theorem}
\label{thm:qi=subs}
If $D$ and $O$ are a faithful encoding of $\sysfnt{I}$ in
$\sysfnt{P}$, then one can identify a subsystem encoding
$\cH_{\sysfnt{P}} = \cH_{\sysfnt{I'}}\tensor\cH_{\sysfnt{S}}\oplus
\cH_{\sysfnt{R}}$ such that for all $\rho$, $D(\rho)$ has support in
$\cH_{\sysfnt{I'}}\tensor\cH_{\sysfnt{S}}$, and for all $A$,
$\cH_{\sysfnt{I'}}\tensor\cH_{\sysfnt{S}}$ and $\cH_{\sysfnt{R}}$ are
invariant subspaces of $O(A)$, and $O(A)$ acts as $A'\tensor\one$ on
$\cH_{\sysfnt{I'}}\tensor\cH_{\sysfnt{S}}$.
\end{Theorem}

The conclusion of the theorem does not hold if we assume only faithful
statics and faithful unitary dynamics.  For example, any irreducible
representation of $U(\cH_{\sysfnt{I}})$ leads to an encoding
satisfying these two faithfulness requirements, and such
representations that have dimension larger than $N$ exist.  For
example, if $\sysfnt{I}$ is a qubit, then any spin $>1/2$
representation of $SU(2)$ yields an encoding that lacks faithful
measurement dynamics.  An other example is ensemble quantum computing
with pure or pseudo-pure
states~\cite{cory:qc1996a,chuang:qc1997a,knill:qc1998c}.  In the case
of pseudo-pure states, faithful statics is only satisfied up to a
scale. Nevertheless, quantum information is still associated with
subsystems.  It may be interesting to determine the nature of
encodings satisfying only faithful statics (perhaps weakened to allow
for an overall scale factor) and faithful unitary dynamics. Are they
always equivalent to a sum of subsystems transforming under distinct
irreducible representations of $U(\cH_{\sysfnt{I}})$?  On the other
hand, we conjecture that faithful statics and measurement dynamics
imply faithful unitary dynamics. However, our proof of the theorem
requires all three faithfulness properties.

An apparently more general faithfulness property, ``faithful
interactions'', requires that the encoding of $\sysfnt{I}$ in
$\sysfnt{P}$ behaves correctly in interactions with other idealized
systems. Faithful interactions are needed if we use the encoded
quantum information in a quantum information processing setting with
multiple physical systems, each encoding quantum information in some
way. Faithful measurement dynamics can be seen to be a special case of
faithful interactions, and, according to the theorem, it implies
faithful interactions in general.

\proofof{~\ref{thm:qi=subs}} Let $\cV$ be the linear sum of the
supports of operators in $D(\rho)$ for all $\rho$. Let $\cV^\perp$ be
its orthogonal complement.  By assumption, $\cV\subseteq
\cH_{\sysfnt{P}}$.  The proof proceeds in three stages. In the first,
we show that the operators of $O(A)$ have $\cV$ and $\cV^\perp$ as
invariant subspaces. We can then redefine $O(A)$ by restricting its
operators to $\cV$. $\cH_{\sysfnt{R}}$ is identified as
$\cV^\perp\cap\cH_{\sysfnt{P}}$.  We then show that $O(A)$ consists of
exactly one operator and deduce that $O$ extends to an algebra
isomorphism when restricted to commuting subsets of observables. The
underlying reason for this involves showing that the eigenspaces of
$O(A)$ may be faithfully identified with eigenspaces of $A$. The
first two stages of the proof do not require faithful unitary
dynamics. The last stage involves analyzing $SU(2)$ subgroups of
$U(\cH_{\sysfnt{I}})$ and corresponding subgroups of $U(\cV)$ induced
by $O$. Their action on eigenspaces of operators in the range of $O$
implies the desired subsystem encoding.

For an operator $X$, let $\Null(X)$ denote the null space of $X$.

\begin{Lemma}
\label{lemma:eigendec}
Let $X\in O(A)$.  Then $\Null(X-\lambda)\cap\cV$ is non-empty if and
only if $\lambda$ is in the spectrum of $A$. Furthermore,
$\cV=\sum_{\lambda}\left(\Null(X-\lambda)\cap\cV\right)$, and
$\Null(X-\lambda)\cap\cV$ is the linear span of the supports of
$\rho\in D(\sigma)$ with $\Supp(\sigma)\subseteq\Null(A-\lambda)$.
\end{Lemma}

\proof Suppose that $\lambda$ is in the spectrum of $A$, and consider
any $\rho\in D(\Pi(A,\lambda))$. By faithfulness of measurement
dynamics, $\Pi(X,\lambda)\rho\Pi(X,\lambda)\in D(\Pi(A,\lambda))$.  By
faithfulness of statics, $\Pi(X,\lambda)\rho\Pi(X,\lambda)$ is not
zero. Since the support of $\Pi(X,\lambda)\rho\Pi(X,\lambda)$ is
contained in $\Null(X-\lambda)\cap\cV$, this intersection is
non-zero. Conversely, suppose that $\Null(X-\lambda)\cap\cV$ is
non-empty. Then there exist $\sigma$ and $\rho\in D(\sigma)$ such that
the support of $\rho$ is not orthogonal to $\Null(X-\lambda)$.  Thus
$\Pi(X,\lambda)\rho\Pi(X,\lambda)$ is not zero and is a member of
$D(\Pi(A,\lambda)\sigma\Pi(A,\lambda))$.  Because
$D(\zero)=\{\zero\}$, $\Pi(A,\lambda)\sigma\Pi(A,\lambda)$ is not
zero. Hence $\Null(A-\lambda)$ is non-zero, so that $\lambda$ is in
the spectrum of $A$.

To prove that $\cV$ is spanned by the subspaces $\Null(X-\lambda)\cap\cV$,
we use the following sequence of inclusions:
\begin{eqnarray}
\cV &\supseteq&
\sum_{\lambda}\left(\Null(X-\lambda)\cap\cV\right) \nonumber\\
   &\supseteq&
     \sum_{\lambda}\sum_\sigma\sum_{\rho:\rho\in D(\sigma)}
        \left(\Supp(\Pi(X,\lambda)\rho\Pi(X,\lambda))\cap\cV\right)\nonumber\\
   &=&
     \sum_\sigma\sum_{\rho:\rho\in D(\sigma)}\sum_{\lambda}
        \left(\Supp(\Pi(X,\lambda)\rho\Pi(X,\lambda))\cap\cV\right)\nonumber\\
   &\supseteq&
     \sum_\sigma\sum_{\rho:\rho\in D(\sigma)}
        \Supp(\rho)\\
   &=& \cV,
\label{eq:incseq}
\end{eqnarray}
where in each expression, $\lambda$ ranges over the spectrum of $A$.
The critical step in the sequence requires the inclusion
\begin{equation}
\sum_{\lambda}
        \left(\Supp(\Pi(X,\lambda)\rho\Pi(X,\lambda))\cap\cV\right)
             \supseteq \Supp(\rho).
\end{equation}
To prove this inclusion, observe that
$\Supp(\Pi(X,\lambda)\rho\Pi(X,\lambda))\subseteq\cV$ because
$\Pi(X,\lambda)\rho\Pi(X,\lambda)\in
D(\Pi(A,\lambda)\rho\Pi(A,\lambda)$. Faithfulness of measurement
dynamics implies that $\Pi(X,\lambda)\rho\Pi(X,\lambda)=\zero$ for
$\lambda$ not in the spectrum of $A$. It then suffices to recall that
for a complete set of orthogonal projectors $P_i$,
$\sum_i\Supp(P_i\rho P_i)=\Supp(\sum_i P_i\rho P_i)\supseteq
\Supp(\rho)$.  \ignore{ This is not entirely obvious, but I think,
well-known.  One way to prove it is to show that if $\ket{x}$ is
orthogonal to the support of $\sum_i P_i\rho P_i$, then it is
orthogonal to the support of $\rho$. To prove this, it helps to
decompose $\rho=\sum_j\ketbra{y_j}$.  Using positivity, one can see
that the first orthogonality statement implies that
$\sum_i\bra{x}P_i\ketbra{y_j}P_i\ket{x}=0$, from which one gets
$\bra{x}P_i\ketbra{y_j}P_i\ket{x}=0$ for each $i$, or $\bra{x}
P_i\ket{y_j}=0$ and because $\sum_iP_i=\one$, $\braket{x}{y_j} = 0$.
}

For the last claim of the lemma, let $W$ be the set of density
operators with support in $\Null(A-\lambda)$.  If $\sigma\in W$, then
$\Pi(A,\lambda)\sigma\Pi(A,\lambda)=\sigma$ and for all
$\lambda'\not=\lambda$, $\Pi(A,\lambda')\sigma\Pi(A,\lambda')=\zero$.
Faithful measurement dynamics imply that for $\rho\in D(\sigma)$,
$\Supp(\rho)\subseteq \Null(X-\lambda)$. Thus, $\sum_{\sigma\in
W}\sum_{\rho\in D(\sigma)}\Supp(\rho) \subseteq
\Null(X-\lambda)\cap\cV$. The following sequence of relationships
proves the reverse inclusion:
\begin{eqnarray}
\Null(X-\lambda)\cap\cV &=&
  \Null(X-\lambda)\cap
   \left(\sum_\sigma\sum_{\rho:\rho\in D(\sigma)} \Supp(\rho)\right)
     \nonumber\\
  &\subseteq&
    \Pi(X,\lambda)
      \left(\sum_\sigma\sum_{\rho:\rho\in D(\sigma)} \Supp(\rho)\right) \nonumber \\
  &=&\sum_\sigma\sum_{\rho:\rho\in D(\sigma)}
     \Supp(\Pi(X,\lambda)\rho\Pi(X,\lambda))\nonumber\\
  &\subseteq&
     \sum_\sigma\sum_{\rho:\rho\in D(\Pi(A,\lambda)\sigma\Pi(A,\lambda))}
     \Supp(\rho)\nonumber\\
  &=&
     \sum_{\sigma\in W}\sum_{\rho:\rho\in D(\sigma)}
     \Supp(\rho),
\end{eqnarray}
where we have used the fact that
for a projector $\Pi$ and a positive semidefinite hermitian operator
$\rho$, $\Pi\Supp(\rho) = \Supp(\Pi\rho\Pi)$.
\qed

\begin{Corollary}
\label{cor:invariant}
Let $X\in O(A)$. Then $\cV$ and $\cV^\perp$ are invariant
subspaces of $X$.
\end{Corollary}

\proof
Because the eigenspaces of $X$ are orthogonal,
Lemma~\ref{lemma:eigendec} implies that $X$ can be block diagonalized
with respect to an orthonormal basis whose first members
span $\cV$.
\qed

Lemma~\ref{lemma:eigendec} and Cor.~\ref{cor:invariant} imply that
without loss of generality, we can assume that for all $X\in O(A)$,
$X$ restricted to $\cV^\perp$ is $\zero$.  If not, replace every
member of $O(A)$ with its restriction to $\cV$. This does not affect
any of the faithfulness requirements.

The last statement in Lemma~\ref{lemma:eigendec} together with the
assumption that $X\in O(A)$ has trivial action on $\cV^\perp$ implies
that $X$'s eigenspaces and eigenvalues are determined by $A$ and the
map $D$. It follows that $O(A)$ consists of exactly one
operator. Thus, without loss of generality, we now take $O(A)$ to be a
function from observables of $\cH_{\sysfnt{I}}$ to observables of
$\cH_{\sysfnt{P}}$.  Lemma~\ref{lemma:eigendec} also implies that
inclusion relationships between eigenspaces of observables of
$\cH_{\sysfnt{I}}$ are preserved by $O$.

\begin{Corollary}
\label{cor:subcommute}
Suppose that $\Null(A-\lambda_1)\subseteq \Null(B-\lambda_2)$.
Then $\Null(O(A)-\lambda_1)\subseteq \Null(O(B)-\lambda_2)$.
\end{Corollary}

If observables $A$ and $B$ commute, we can construct an observable $C$
whose eigenspaces are the maximal common eigenspaces of $A$ and $B$.  By
using the eigenspaces of $O(C)$ to derive the eigenspaces of $O(A)$ and
of $O(B)$, we can see that $O(AB) = O(A)O(B)$ and $O(\alpha A+\beta B) =
O(\alpha A)+O(\beta B)$, so that $O$ preserves the algebraic structure
of commuting sets of observables. Similarly,
for any eigenbasis $\ket{\lambda_i}$ of $A$, we can use
an operator $C$ with non-degenerate eigenvalues having the same
eigenbasis to see that the spaces
$\Supp(D(\ket{\lambda_i}\bra{\lambda_i}))$
are a complete orthogonal decomposition of $\cV$ into eigenspaces
of $A$. From this it follows that $O(A)$ is determined
by the values of $D$ on pure states.

For the last stage of the proof of Thm.~\ref{thm:qi=subs}, we fix an
orthonormal basis $\kets{i}{I}$ of $\cH_{\sysfnt{I}}$. Let $e_{ij} =
\kets{i}{I}\bras{j}{I}$, $X_{ij} = e_{ij}+e_{ji}$, $Y_{ij} = -i
e_{ij}+ie_{ji}$, $C=\sum_{i} i e_{ii}$ and $\cV_i = \Null(O(C)-i)$.
Note that $ie^{-i X_{ij}\pi/2} \kets{i}{I} = e^{-i
Y_{ij}\pi/2}\kets{i}{I}= \kets{j}{I}$. According to faithfulness of
unitary dynamics, $e^{-iO(A)}D(\sigma)e^{iO(A)} \subseteq
D(e^{-iA}\sigma e^{iA})$. (For sets $D$ and operators $U$, $UD =
\{Ux\suchthat x\in D\}$.)  The inclusion is an equality because we
also have $D(e^{-iA}\sigma e^{iA}) =
e^{-iO(A)}e^{iO(A)}D(e^{-iA}\sigma e^{iA})e^{-iO(A)} e^{iO(A)}
\subseteq e^{-iO(A)} D(\sigma ) e^{iO(A)}$.  This and the earlier
results imply that $e^{-iO(X_{ij})\pi/2} \cV_i = \cV_j$.  By using
Cor.~\ref{cor:subcommute} with the eigenspaces of $X_{ij}$, $Y_{ij}$
and $e_{ii}+e_{jj}$, we can see that the non-zero eigenspaces of
$O(X_{ij})$ and $O(Y_{ij})$ are contained in $\cV_i\oplus \cV_j$.

Because of the algebraic properties of $O$ mentioned above and
$X_{ij}^2=e_{ii}+e_{jj}$, $e^{-iO(X_{ij})\pi/2} = -iO(X_{ij})$.  We
can therefore fix an orthonormal basis $\ket{il}$ of $\cV_i$ such that
$O(X_{0i})\ket{0l} = \ket{il}$ and $O(X_{0i})\ket{il}=\ket{0l}$.  The
goal is to show that we can identify $\ket{il}$ with
$\kets{i}{I'}\kets{l}{S}$ such that $O(A)$ acts as the identity on the
cosubsystem $\sysfnt{S}$.  Note that the operators $X_{0j}$ and
$e_{jj}$ generate the Lie algebra of $U(\cH_{\sysfnt{I}})$.  Thus
compositions of exponentials of the form $e^{-i X_{0j}t}$ or $e^{-i
e_{jj} s}$ act transitively on the pure states of
$\cH_{\sysfnt{I}}$. It follows that for any $\kets{\psi}{I}$,
$\Supp(D(\kets{\psi}{I}\bras{\psi}{I}))$ is an image of corresponding
compositions of exponentials of the form $e^{-i O(X_{0i}) t}$ or
$e^{-i e_{jj} s}$ acting on $\cV_0$. Such compositions are completely
determined by the basis $\ket{il}$. Now $O(A)$ is determined by
$\Supp(D(\kets{\psi}{I}\bras{\psi}{I}))$, with $\kets{\psi}{I}$
ranging over eigenvectors of $O(A)$.  Since for fixed $l$, $O(X_{0i})$
and $O(e_{jj})$ act as they should on the states $\ket{kl}$, we have
that $O(A)$ necessarily satisfies $\bra{kl}O(A)\ket{kl} =
\bras{k}{I}A\kets{k}{I}$, as desired.  \qed

We return to the issue of the relationship between decoding operations
and faithful encodings. Decoding as defined in the introduction is the
traditional way of identifying quantum information and generalizes
recovery operations. A general form of the situation addressed by
decoding involves one or more encoding isometries
$C_i:\cH_{\sysfnt{I}}\rightarrow\cH_{\sysfnt{P}}$, one or more
possible events $E_i$ that are operators on $\cH_{\sysfnt{P}}$ that
may occur before we decode, and a decoding operation $D$ that (after
purification, if necessary) isometrically maps $\cH_{\sysfnt{P}}$ into
$\cH_{\sysfnt{I}}\tensor\cH_{\sysfnt{A}}$ for some possibly composite
system $\sysfnt{A}$. We say that $\{C_i\}, \{E_j\}, D$ preserve
quantum information if for all $i,j$, $DE_jC_i\kets{\psi}{I} =
\kets{\psi}{I}\kets{\phi_{ij}}{A}$ for some unnormalized vector
$\kets{\phi_{ij}}{A}$ that does not depend on $\kets{\psi}{I}$.  Here,
which event $E_i$ occurred is assumed to be unknown.  We could
consider the case where the decoding operation is chosen after the
events and depends on partial knowledge of the events. However, by
conditioning on the knowledge, we return to the situation just
described. The only difference is that the subsystem associated with
the situation may depend on the partial knowledge.  To capture the
case where quantum information is stored in error-correcting
subsystems, let $C_{i}$ be given by the isometries identifying
$\cH_{\sysfnt{I}}$ with $\cH_{\sysfnt{I'}}\tensor\kets{i}{S}$, where
the $\kets{i}{S}$ range over any spanning set of $\cH_{\sysfnt{S}}$.

In order to justify the subsystems principle, we prove the next theorem.

\begin{Theorem}
\label{thm:ced}
If $\{C_i\}, \{E_j\}, D$ preserve quantum information, then there
exists a subsystem encoding
$\cH_{\sysfnt{P}}=\cH_{\sysfnt{I'}}\tensor\cH_{\sysfnt{S'}}\oplus\cH_{\sysfnt{R}}$
such that for all $i,j$ and $\kets{\psi}{I}$, $E_jC_i\kets{\psi}{I} =
\kets{\psi}{I'}\kets{\phi'_{ij}}{S'}\in\cH_{\sysfnt{P}}$ and
$D\kets{\psi}{I'}\kets{\phi'_{ij}}{S'}=\kets{\psi}{I}\kets{\phi_{ij}}{A}$,
where the $\kets{\phi'_{ij}}{S'}$ and $\kets{\phi_{ij}}{A}$ do not
depend on $\kets{\psi}{I}$.
\end{Theorem}

\proof This follows from the fact that there exists a protectable
subsystem associated with any quantum error-correcting code and
associated recovery operation, which was proven
in~\cite{knill:qc1995e,knill:qc1999b}.  Alternatively, we could prove
the theorem from Thm.~\ref{thm:qi=subs} by defining $D(\kets{\psi}{I})
= \{E_jC_i\kets{\psi}{I}\}$ and $O(A)$ by pullback of the appropriate
operators via the decoding operator $D$. Here we give a direct proof.
Let the $\kets{\phi_{ij}}{A}$ be as required according to the
definition of preserving quantum information. Let $\cS$ be the set of
vectors $\kets{\phi}{A}$ such that
$\cH_{\sysfnt{I}}\tensor\kets{\phi}{A}$ is contained in the range of
$D$. Then $\cS$ contains the $\kets{\phi_{ij}}{A}$.  Because the range
of $D$ is linearly closed, so is $\cS$.  Define
$\cH_{\sysfnt{S}}=\cS$.  Using the isometric properties of $D$, we can
define a subsystem encoding
$\cH_{\sysfnt{P}}=\cH_{\sysfnt{I'}}\tensor\cH_{S'}\oplus\cH_{\sysfnt{R}}$
such that $D(\kets{\psi}{I'}\kets{\phi}{S'}) =
\kets{\psi}{I}\kets{\phi}{S}$. This subsystem encoding has the desired
properties.  \qed

\section{Finding Noiseless Subsystems}
\label{sect:ns}

If $\cA$ is a $\dagger$-closed subalgebra of $B(\cH_{\sysfnt{P}})$,
the canonical decomposition of $\cH_{\sysfnt{P}}$ is
\begin{equation}
\cH_{\sysfnt{P}} = \sum_i
  \cH_{\sysfnt{I_i}}\tensor\cH_{\sysfnt{S_i}}\oplus\cH_{\sysfnt{R}},
\end{equation}
where operators $A\in\cA$ act as $\sum_i
\slb{\one}{I_i}\tensor\slb{S_i(A)}{S_i} + \slb{\zero}{R}$. For every
operator of the form $\sum_i \slb{\one}{I_i}\tensor\slb{B_i}{S_i} +
\slb{\zero}{R}$, there exists an $A\in\cA$ with $S_i(A)=B_i$.  The
$\cH_{\sysfnt{I_i}}$ are noiseless subsystems for $\cA$.  We also
consider $\cH_{\sysfnt{R}}$ to be noiseless for $\cA$, but note that
error operators in $\cA$ have probability zero for states in this
subspace.  The tensor products and direct sums in the decomposition
must be consistent with the Hilbert space's inner product. This is
implicit in the construction and the identification via an isometry.

Let $\cE$ be a linearly closed set of error operators in
$B(\cH_{\sysfnt{P}})$.  For now, we do not assume that $\cE$ is the
span of the operation elements of a quantum operation. Let
$\cH_{\sysfnt{I'}}\tensor\cH_{\sysfnt{S}}\oplus\cH_{\sysfnt{R}}$
define a subsystem encoding of $\sysfnt{I}$ in $\sysfnt{P}$.  Let
$\Pi$ be the projector onto the support of $\sysfnt{I'}$,
$\cH_{\sysfnt{I'}}\tensor\cH_{\sysfnt{S}}\subseteq\cH_{\sysfnt{P}}$.
The subsystem is noiseless for $\cE$ if and only if for all $E\in\cE$,
the restriction of $E$ to $\cH_{\sysfnt{I'}}\tensor\cH_{\sysfnt{S}}$
acts as the identity on $\cH_{\sysfnt{I}}$. Equivalently, for all
$E\in\cE$, $E\Pi = \slb{\one}{I}\tensor \slb{S(E)}{S}$. It is
straightforward to verify that if the subsystem is noiseless, then
$\Pi$ projects onto an invariant subspace of $\cE$ and $\cE\Pi$
generates a $\dagger$-closed subalgebra of operators acting on the
support of $\Pi$ whose canonical decomposition contains noiseless
subsystems with state space dimension at least $N$, the dimension of
$\cH_{\sysfnt{I}}$. Such noiseless subsystems are also noiseless for
$\cE$.  This leads to a strategy for finding noiseless subsystems with
maximum dimensional $\cH_{\sysfnt{I'}}$ that is equivalent to the
strategy proposed in~\cite{choi:qc2005b}: 1. Pick an invariant
subspace of $\cE$ and let $\Pi$ be its projector. 2. Determine the
canonical decomposition of the $\dagger$-closed algebra generated by
$\Pi\cE$. The noiseless subsystems of this algebra are candidate
noiseless subsystems for $\cE$. Our goal is to provide an explicit
algorithm for finding suitable $\Pi$ and associated subsystems. The
algorithm involves the decomposition of a matrix algebra, for which
efficient algorithms are known, as we explain below.  Note that in
addition to the noiseless subsystems identified in this way, one can
construct other noiseless subsystems as subsystems of already obtained
noiseless subsystems, or by combining cosubsystems of identical
dimensional noiseless subsystems with orthogonal supports. These
constructions cannot yield larger dimensional noiseless subsystems,
but they may generate ones with greater error tolerance or more
efficiently controllable states.

Let $\cA$ be the algebra generated by $\cE$. Any noiseless
subsystem for $\cE$ is a noiseless subsystem for $\cA$. $\cA$ is not
necessarily $\dagger$-closed. As a result, $\cA$ does not have a
canonical decomposition of $\cH_{\sysfnt{P}}$ as a direct sum of
tensor products of Hilbert spaces. Nevertheless, we can identify a
special subspace within which a similar decomposition is possible and
where maximum dimensional noiseless subsystems may be found.  This
subspace is the span $\cS$ of the irreducible subspaces of $\cA$.  A
subspace $\cV$ of $\cH_{\sysfnt{P}}$ is ``irreducible'' for $\cA$ if
it is $\cA$-invariant, $\cA\cV\not=0$ and there is no non-zero
$\cA$-invariant proper subspace of $\cV$. Let $\cZ$ be the null space
of $\cA$. Both $\cS$ and $\cZ$ are invariant.

\begin{Lemma}
\label{lemma:maxnsloc}
A maximum dimensional noiseless subsystem for $\cA$ can
be found in $\cS$ or in $\cZ$.
\end{Lemma}

Note that $\cZ$ is itself a noiseless subsystem. This subsystem is
trivial in the sense that the probability of $\cE$-errors is zero for
any state in $\cZ$. This means that in a realistic setting, there must
be operators acting on the system not included in $\cE$, and for $\cZ$
to be at least approximately noiseless, they need to act as
operators close to the identity when restricted to $\cZ$.

\proofofx{Lemma~\ref{lemma:maxnsloc}} Suppose that $\cH_{\sysfnt{I'}}$
is a noiseless subsystem of $\cH_{\sysfnt{P}}$ with cosubsystem
$\cH_{\sysfnt{S}}$.  Then
$\cV=\cH_{\sysfnt{I'}}\tensor\cH_{\sysfnt{S}}$ is invariant under $\cA$
and for $A\in\cA$, $A$ acts as $\slb{\one}{I}\tensor \slb{S(A)}{S}$ on
$\cV$.  If for all $A\in\cA$, $\slb{S(A)}{S}=0$, then $\cV\subseteq\cZ$
and we are done. If not, then there exists a nontrivial irreducible
subspace $\cS_i$ of $\cH_{\sysfnt{S}}$ under the action of
$\{S(A)\suchthat A\in\cA\}$.  For each state $\kets{\psi}{I'}$,
$\kets{\psi}{I'}\tensor\cS_i$ is an irreducible representation for $\cA$. In particular,
$\cH_{\sysfnt{I'}}\tensor\cS_i\subseteq \cS$. Since
$\cH_{\sysfnt{I'}}\tensor\cS_i$ is also a noiseless subsystem, the
proof is complete.  \qed

According to the theory of $R$-modules, $\cS$ is a module for
$\cA$. The definition implies that it is semisimple, from which it
follows that $\cS=\sum_i \cS_i$ where the $\cS_i$ are irreducible and
the sum is over independent subspaces, see~\cite{hungerford:qc1980a},
Chapter 9. $\cS_i$ and $\cS_j$ are isomorphic with respect to the
action of $\cA$ if there exists an invertible linear map $U_{ij}$ from
$\cS_i$ to $\cS_j$ such that for $\ket{x}\in\cS_i$, $AU_{ij}\ket{x} =
U_{ij}A\ket{x}$. The map $U_{ij}$ is said to ``intertwine'' $\cS_i$
and $\cS_j$.  We can relabel the $\cS_i$ to form sets $\{\cS_{ik}\}_i$ of
isomorphic irreducible representations. For each $k$, let $\cV_k$ be the span of the
$\cS_{ik}$ and let $U^{(k)}_{0j}$ be an intertwiner from $\cS_{0k}$ to
$\cS_{jk}$. Choose a basis $\ket{i0k}$ of $\cS_{0k}$ and define
$\ket{ijk} = U^{(k)}_{0j}\ket{i0k}$.  Note that these vectors need not
be orthogonal or normalized.  Nevertheless, they define invertible
linear maps from tensor products $\cJ_k\tensor\cS_{0k}$ to $\cV_k$ via
the linear extension of $\ket{j}\tensor\ket{i0k}\mapsto\ket{ijk}$.
The action of $A\in\cA$ with respect to this factorization is on
$\cS_{0k}$ only.

\begin{Lemma}
\label{lemma:maxnslock}
A maximum dimensional noiseless subsystem for $\cA$ in $\cS$
can be found in one of the $\cV_k$.
\end{Lemma}

\proof This follows from the argument given in the proof of
Lemma~\ref{lemma:maxnsloc}. It suffices to observe that the
irreducible representations $\kets{\psi}{I}\tensor\cS_i$ are
isomorphic for different $\kets{\psi}{I}$. \qed

The main remaining problem in narrowing the search space for maximum
dimensional noiseless subsystems is that the factorization of the
$\cV_k$ may fail to preserve the inner product. To simplify the
notation, fix $k$ and let $\cV=\cV_k$, $\cS_0=\cS_{0k}$ and
$\cJ=\cJ_k$.  Let $U$ be an invertible linear map from
$\cJ\tensor\cS_0$ to $\cV$ that implements the above-mentioned
factorization of $\cV$.  Thus, for $A\in\cA$ and
$\ket{x}\in\cJ\tensor\cS_0$, $AU\ket{x} = U(\one\tensor R(A))\ket{x}$,
where $R$ is a well-defined, irreducible representation of $\cA$ on
$\cS_0$. Note that an irreducible representation of $\cA$ on $\cS_0$
is onto $B(\cS_0)$ (Burnside's theorem). This implies that any
noiseless subsystem of $\cV$ must be associated with a subspace $\cJ'$
of $\cJ$ such that the restriction of $U$ to $\cJ'\tensor\cS_0$ has
the property that there are linear operators $W$ on $\cJ'$ and $V$ on
$\cS_0$ such that $U (W\tensor V)$ is an isometry.  Fortunately, in
cases where $\cA$ is generated by the operational elements of a
quantum operation, we do not need to search for such subspaces.

\begin{Lemma}
\label{lemma:nsunit}
If $\cA$ is generated by the operational elements of a quantum
operation, then there exist linear operators $W$ on $\cJ$ and $V$ on
$\cS_0$ such that $U(W\tensor V)$ is unitary.
\end{Lemma}

\proof Let $\{E_i\}_i$ generate $\cA$, where the $E_i$ are the
operational elements of a quantum operation $\cO$. By composing $\cO$
with itself sufficiently many times, it is possible to obtain a
quantum operation $\cO'$ such that its operational elements span
$\cA$. Thus, without loss of generality, assume that the $E_i$ span
$\cA$ and $\sum_i E_i^\dagger E_i = \one$.  We have $E_i =
U(\one\tensor R(E_i))U^{-1}$.  In order to continue, assume, without
loss of generality, that $\cJ\tensor\cS_0 =\cV$. This can be done by
means of any isometry between $\cV$ and $\cJ\tensor\cS_0$.  This
implies that $U$ is an invertible but not necessarily unitary linear
map from $\cJ\tensor\cS_0$ to itself.  We have
\begin{equation}
\sum_i {U^{-1}}^\dagger (\one \tensor R(E_i)^\dagger) U^\dagger U
          (\one\tensor R(E_i)) U^{-1} = \one,
\end{equation}
or, equivalently,
\begin{equation}
\sum_i (\one\tensor R(E_i)^\dagger ) U^\dagger U
          (\one\tensor R(E_i)) = U^\dagger U.
\end{equation}
This implies that for all positive semidefinite $\sigma$
on $\cJ$, 
\begin{equation}
\label{eq:fixed}
\sum_i R(E_i)^\dagger \trace_\cJ((\sigma\tensor\one)U^\dagger U (\sigma\tensor\one)) R(E_i)
 = \trace_\cJ((\sigma\tensor\one)U^\dagger U (\sigma\tensor\one)),
\end{equation}
where $\trace_\cJ$ is the partial trace over $\cJ$.  Let $\mathbf{R}$
be the operation defined by $\mathbf{R}(X) = \sum_i R(E_i)^\dagger X
R(E_i)$.  According to Eq.~\ref{eq:fixed}, for all positive
semidefinite $\sigma$, $\trace_\cJ((\sigma\tensor\one)U^\dagger
U(\sigma\tensor\one))$ is a positive semidefinite fixed point of $R$.
The spanning assumption on the $E_i$ and irreducibility of $\cS_0$
under $R(\cA)$ imply that the $R(E_i)$ span $B(\cS_0)$.  It follows
that if $\rho\not=\zero$ is positive semidefinite and
$\mathbf{R}(\rho) = \rho$, then the support of $\rho$ is $\cS_0$. It
also implies that $\mathbf{R}$ has at most one positive
fixed point (up to positive multiples): If $\rho'$ is
another one, then so is $\rho-\epsilon\rho'$ for all $\epsilon$.  Let
$\epsilon$ be the largest such that $\rho-\epsilon\rho'$ is positive
semidefinite. Then $\rho-\epsilon\rho'$ is a fixed point with
non-maximal support, which implies that it is $\zero$.  Let $\rho$ be
the unique trace $1$ positive fixed point of $R$. Then, for all
positive semidefinite $\sigma$,
$\trace_\cJ((\sigma\tensor\one)U^\dagger U (\sigma\tensor\one))$ is a
multiple of $\rho$.  We can now deduce that $U^\dagger U =
\rho'\tensor\rho$ for some strictly positive $\rho'$.  \ignore{ This
is not entirely trivial. But we can see that for all $X$ orthogonal to
$\rho$ in the trace inner product,
$\trace((\sigma\tensor\one)U^\dagger U (\sigma\tensor\one)
(\one\tensor X)) = \trace(\rho X) = 0$ The left hand side is $\trace(
U^\dagger U (\sigma^2\tensor X))$, by cyclicity of trace. The result
then follows because the operators $\sigma^2$ for positive
semidefinite $\sigma$ linearly span all operators on $\cJ$. } Defining
$V=\rho^{-1/2}$ and $W=\rho'^{-1/2}$ yields the lemma.  \qed

The above suggests the following strategy for finding
maximum-dimensional noiseless subsystems: 1. Determine the span $\cS$
of the irreducible subspaces of $\cA$. 2.  Decompose $\cS$ into a
direct sum $\bigoplus_i \cI_i$ of subspaces spanned by isomorphic
irreducible subspaces. 3. For each $\cI_i$, let $\cA_i$ be the
restriction of $\cA$ to $\cI_i$ and find the canonical decomposition
for the $\dagger$-closed algebra generated by $\cA_i$.  This strategy
will find maximum-dimensional noiseless subsystems provided that $\cA$
is generated by the operational elements of a quantum operation.
There are efficient algorithms for each step of this strategy; for a
review, see~\cite{struble:qc2000a}. For completeness, we outline an
algorithm that implements the strategy. 

To find $\cS$, consider the structure of $\cA$ in more detail.  If
$\cA$ does not contain $\one$, replace $\cA$ by $\cA+\cmplx\one$.  By
doing so, the action of $\cA$ on $\cZ$ is no longer zero, but $\cZ$ is
still distinguishable from the other irreducible subspaces. Every
one-dimensional subspace of $\cZ$ is irreducible and not isomorphic to
the irreducible subspaces of $\cS$.  There exists a maximal chain of
invariant subspaces
$0=\cV_0\subset\cV_1\subset\ldots\subset\cV_n=\cH_{\sysfnt{P}}$ such
that the action of $\cA$ induced on the quotients $\cV_{k+1}/\cV_k$ is
irreducible or zero. In a basis $\ket{e_{kj}}$ of $\cH_{\sysfnt{P}}$
where $\ket{e_{(k+1)j}}\in\cV_{k+1}\setminus\cV_{k}$ ($\setminus$
denotes set difference), the operators of $\cA$ are block upper
triangular.  Let $\cJ$ be the members of $\cA$ that act as $\zero$ on
each of these quotients. $\cJ$ is known as the Jacobson radical of
$\cA$. Let $\cN$ be the null space of $\cJ$, which is the set of
vectors in the intersection of the null spaces of operators of
$\cJ$. Then $\cN$ is invariant (because $\cJ$ is a two-sided ideal)
and $\cS\subseteq \cN$ (because $\cS$ is invariant and the span of
irreducible subspaces).  A fundamental property of $\cJ$ is that
$\cA/\cJ$ is a semisimple algebra. Let $\cA_\cN$ be the restriction of
$\cA$ to $\cN$.  Then $\cA_\cN$ is isomorphic to a quotient of
$\cA/\cJ$, which implies that $\cA_\cN$ is semisimple.  According to
the representation theory of semisimple algebras, $\cN$ is a
semisimple $\cA_\cN$-module, which implies that $\cN=\cS$. Thus, to
determine $\cS$, we can use an efficient algorithm for finding the
Jacobson radical and then compute its null space. Decomposing $\cN$
into independent irreducible subspaces can be done by means of an
efficient algorithm for the decomposition of semisimple algebras over
the complex numbers. A randomized algorithm can be based on the
observation that $\cA_\cN$ is isomorphic to a direct sum of complete
matrix algebras $\cA_k$ on the $\cS_{0k}$, acting canonically on the
(non-unitary) decomposition of $\cS$ into a direct sum of products
$\cI_k=\cJ_k\tensor\cS_{0k}$. It follows that a random matrix in
$\cA_\cN$ (with respect to a suitably chosen probability distribution)
typically has generalized eigenspaces that generate (by
multiplication by members of $\cA_\cN$) exactly one of the invariant
subspaces $\cI_k=\cJ_k\tensor\cS_{0k}$. This yields the desired matrix
algebras $\cA_k$. For each $\cA_k$, let $\cA^*_k$ be the
$\dagger$-closed algebra generated by $\cA_k$. The canonical
factorization of $\cI_k$ with respect to $\cA^*_k$ can also be
obtained by a randomized algorithm. By construction,
$\cI_k=\cH_1\tensor\cH_2$ (isometrically), with $\cA^*_k$ acting only
on $\cH_2$.  The eigenspaces of a randomly chosen Hermitian operator
$H_2$ in $\cA^*_k$ are typically of the form $\cH_{1i}\tensor\ket{i}$
for an orthonormal basis of $\cH_2$, where $\cH_{1i}=\cH_1$, but with
the isometry for making this identification not yet known.  With high
probability, these isometries can be determined from an independently
chosen second $H'_2$ by expressing $H'_2$ in an orthonormal basis
whose $i$'th block of vectors is a basis of
$\cH_{1i}\tensor\ket{i}$. Because $H'_2$ is a Kronecker product with
identity action on $\cH_1$, the $i,j$ block of $H'_2$ must define an
isometry between $\cH_{1i}$ and $\cH_{1j}$ (if it is nonzero). These
isometries must be consistent and induce the desired tensor product
structure.

Components of the algorithm of the previous paragraph not given
explicitly include the generation of an algebra from a set of matrices
(this comes up in generating $\cA$ from an error set and generating
$\dagger$-closed algebras from a given one) and various standard matrix
manipulations such as matrix multiplication, eigenvalue and eigenspace
determination, etc. We do not discuss the latter here.  To generate
the matrix algebra from a set of operators $\{E_i\}$, assume without
loss of generality that the $E_i$ are independent. Then iteratively,
choose $i,j$ and determine whether $E_i E_j$ is in the span of the
$E_i$. If not, adjoin it to the set. Stop when for all $i,j$, $E_i
E_j$ is in the span of the $E_i$.

\section{Protectable Subsystems}
\label{sect:initialization}

As above, let $\{E_i\}_i$ be a set of error operators on
$\cH_{\sysfnt{P}}$. Let $\cH_{\sysfnt{P}} =
\cH_{\sysfnt{I'}}\tensor\cH_{\sysfnt{S}}\oplus\cH_{\sysfnt{R}}$ be a
subsystem encoding. The subsystem $\sysfnt{I'}$ is ``initialization
protectable'' (or ``protectable'' for short), if there exists a
quantum operation with operation elements $\{R_i\}_i$ such that
$\sysfnt{I'}$ is noiseless for $\{E_iR_j\}_{i,j}$.  The goal of this
section is to reduce the problem of determining whether a given
subsystem is protectable to the problem of searching for certain
extremal error-detecting codes. We then reduce this problem to several
linear algebra problems.

Let $\kets{i}{S}$ be an orthonormal basis of $\cH_{\sysfnt{S}}$.  For
any state $\ket{\psi}$ of
$\cH_{\sysfnt{I'}}\tensor\cH_{\sysfnt{S}}\subseteq\cH_{\sysfnt{P}}$, we
define $\bras{i}{S}\ket{\psi}\in\cH_{\sysfnt{I'}}$ by the identity
$\sum_i(\bras{i}{S}\ket{\psi})\tensor\kets{i}{S}=\ket{\psi}$.  Let
$\cV$ be the intersection of the inverse images of
$\cH_{\sysfnt{I'}}\tensor\cH_{\sysfnt{S}}$ under the errors $E_i$.

\begin{Lemma}
\label{lemma:protiso}
With the definitions of the previous paragraphs,
$\sysfnt{I'}$ is protectable if and only if there exists a
subspace $\cD\subseteq\cV$ with the property that
the maps $F_{ij}:\ket{\psi}\mapsto \bras{j}{S}E_i\ket{\psi}$
are proportional to a single isometry from $\cD$ to $\cH_{\sysfnt{I'}}$.
\end{Lemma}

\proof For the ``if'' part of the lemma, we show that $\cD$ is an
error-correcting code for $\{E_i\}_i$.  We can reconstruct $E_i$ on
$\cD$ from the $F_{ij}$ by the identity $E_i\ket{\psi} = \sum_j
(F_{ij}\ket{\psi})\tensor\kets{j}{S}$.  Let $U$ be the isometry such
that $\bras{j}{S}E_i\ket{\psi} = \alpha_{ij} U\ket{\psi}$.  Then
$E_i\ket{\psi} = (U\ket{\psi})\tensor \sum_j\alpha_{ij}\kets{j}{S}$.
That $\cD$ is an error-correcting code follows immediately.  The
operators $R_i$ are given by $U^{-1}\ketbras{i}{i}{S}$.

For the converse, we can use the subsystems principle (more
specifically, Thm.~\ref{thm:ced}), according to which there must be a
subsystem decomposition $\cH_{\sysfnt{P}} =
\cH_{\sysfnt{I''}}\tensor\cH_{\sysfnt{T}}\oplus \cH_{\sysfnt{Q}}$ such
that $R_i(\kets{\psi}{I'}\tensor\kets{j}{S}) =
\kets{\psi}{I''}\tensor\kets{\phi_{ij}}{T}$ and
$E_i\kets{\psi}{I''}\kets{j}{T}=\kets{\psi}{I'}\kets{\varphi_{ij}}{S}$.
The desired subspace is given by
$\cH_{\sysfnt{I''}}\tensor\kets{0}{T}$ for any base state $\kets{0}{T}$
of $\sysfnt{T}$. Note that the desired isometry is implicity defined
via the two subsystem decompositions.  \qed

The maps $F_{ij}$ defined in the statement of
Lemma~\ref{lemma:protiso} are well defined from $\cV$ to
$\cH_{\sysfnt{I}}$.  Let $M,N$ be the dimensions of $\cV$ and
$\cH_{\sysfnt{I}}$, respectively.  By choosing orthonormal bases
$\{\kets{i}{V}\}_i$ of $\cV$ and $\{\kets{j}{I'}\}_j$ of
$\cH_{\sysfnt{I'}}$, the $F_{ij}$ are expressible as $N\times M$
matrices (also denoted by $F_{ij}$) with entries
$(F_{ij})_{kl}=\bras{k}{I'}F_{ij}\kets{l}{V}$.  Without loss of
generality, $M\geq N$, for otherwise the subsystem $\sysfnt{I'}$ is
clearly not protectable.  The condition in Lemma~\ref{lemma:protiso}
can be seen to be equivalent to the requirement that there exists a
unitary matrix $V$ such that $F_{ij} V$ contains a multiple of the
$N\times N$ identity matrix as its first $N\times N$ block. The code
$\cD$ is spanned by the first $N$ columns of $V$.  This requirement is
reminiscent of the familiar condition on the existence of an
$N$-dimensional error-detecting quantum code, according to which there
must exist a unitary matrix $W$ such that $W E_i W^\dagger$ has a
multiple of an $N\times N$ identity matrix as its first diagonal
subblock. The protectability requirement can indeed be reduced to the
existence of an error-detecting code. In particular, $\sysfnt{I'}$ is
protectable if and only if there exists an $N$-dimensional
error-detecting code for the operators $\{F_{ij}^\dagger F_{kl}\}$.
Note that this is equivalent to requiring the existence of an
$N$-dimensional error-correcting code for the operators $F'_{ij}$,
where $F'_{ij}$ is the square matrix obtained from $F_{ij}$ by
expanding with rows of zeros.  However, we do not have to consider all
operators $F_{ij}^\dagger F_{kl}$. It suffices to find an
$N$-dimensional error-detecting code for operators of the form
$F_{ij}^\dagger F_{ij}$ and $F_{ij}^\dagger F_{\pi(ij)}$, where $\pi$
is a cyclic permutation of the index pairs.

We call subspaces $\cD$ satisfying the condition in
Lemma~\ref{lemma:protiso} ``protecting'' codes (for $\sysfnt{I'}$).
There are several procedures that can be used to reduce the difficulty
of the search for protecting codes.

\begin{Lemma}
\label{lemma:protnull}
All protecting codes are contained in the null space of the $F$ in
the linear span of $\{F_{ij}\}_{ij}$ whose rank is strictly less than $N$.
\end{Lemma}

\proof Let $V$ be as specified in the paragraph before the statement
of the lemma.  If the rank of $F$ is less than $N$, then the first
$N\times N$ block of $FV$ must be zero, from which the result follows.
\qed

Let $G_1,\ldots, G_k$ be $N\times M$ matrices.  We say that
$\{G_1,\ldots G_k\}$ has maximal row rank if the span of the rows of
the $G_i$ has dimension $kN$. The next lemma generalizes
Lemma~\ref{lemma:protnull}.

\begin{Lemma}
\label{lemma:protextnull}
Let $G_1,\ldots,G_k$ be in the span of the $F_{ij}$ such that
$\{G_1,\ldots, G_k\}$ does not have maximal row rank, but for every
$k-1$ independent $G'_1,\ldots G'_{k-1}$ in the span of the $G_l$,
$\{G'_1,\ldots G'_{k-1}\}$ has maximal row rank.  Then any protecting
codes are contained in the intersection of the null spaces of the
$G_i$.
\end{Lemma}

\proof Let $V$ be such that $G_iV$ has an initial block proportional
to the $N\times N$ identity matrix and $\cD$ is spanned by the first
$N$ columns of $V$.  Suppose that $\cD$ is not contained in the
null space of some $G_i$. Then $G_iV$'s initial $N\times N$ block is
not zero. The space $\cG$ of matrices $G$ in the span of the $G_j$
such that $GV$ has an initial $N\times N$ zero block is
$k-1$-dimensional.  Because the row span of $G_i$ is independent of
the linear span $\cR$ of the rows of the matrices in $\cG$, the
dimension of $\cR$ is strictly less than $(k-1)N$, contradicting the
assumption of the lemma.  \qed

Lemma~\ref{lemma:protextnull} means that in principle, the problem of
finding $\cD$ can be reduced to the case where each $F_{ij}$ has full
rank and its row space is independent of the space spanned by the rows
of the other $F_{kl}$.  In this case there are at most $M/N$
independent matrices $F_{ij}$.  Unfortunately, we do not know of an
efficient algorithm for checking the condition of
Lemma~\ref{lemma:protextnull} that would enable reducing the problem
to this case. Nevertheless, we can show that one can reduce to the
case where there are at most $M-1$ independent $F_{ij}$.

\begin{Lemma}
\label{lemma:protextM}
For $N>1$, if there are $M$ or more independent $F_{ij}$,
then there exists a nonzero $G$ in the span of the $F_{ij}$ such
that $G$ does not have full rank.
\end{Lemma}

\proof Let $\{G_i\}_{i=1}^{l}$ be a basis of the linear span of the
$F_{ij}$.  Let $g^j_{i}$ be the $j$'th row of $G_i$.  If one of the
$g^j_{i}$ is zero, we are done.  Suppose $l>M$. Then the $g^1_{i}$ are
dependent, so there is a non-trivial linear combination of the $G_i$
with zero first row.  Suppose $l=M$. Consider the matrices $A_j$ whose
$i$'th rows are the $g^j_{i}$. Then there exists a non-zero linear
combination $\alpha A_1+\beta A_2$ with determinant zero. Let
$x\not=\mathbf{0}$ be in the null space of $(\alpha A_1+\beta A_2)^T$.
Then $G=\sum_ix_iG_i$ is not zero and the row vector
$y=(\alpha,\beta,0\ldots)$ satisfies $yG=0$, so that $G$ does not have
full rank.  \qed

Note that the proof of the lemma contains an efficient algorithm
for finding a non-full rank $G$.

Let $\slb{\rho}{AB}$ be a density matrix on systems $A$ and $B$.  What
states $\sigma$ of $\sysfnt{A}$ can be obtained by projecting
$\sysfnt{B}$ onto a pure state $\kets{\psi}{B}$?  The ``pure
$\sigma$-projection problem'' for $\slb{\rho}{AB}$ is to determine a
state $\kets{\psi}{B}$ such that
$\bras{\psi}{B}\slb{\rho}{AB}\kets{\psi}{B} = p\sigma$ for some $p\not=0$,
if such a state exists.

\begin{Theorem}
\label{thm:projection}
If the span of the rows of the $F_{ij}$ is $M$-dimensional, the
problem of determining whether subsystem $\sysfnt{I'}$ is protectable
is efficiently reducible to a pure $\one$-projection problem.
\end{Theorem}

If the rows of the $F_{ij}$ do not span the full space, then the
protectability problem may be reduced to a generalization of the pure
$\one$-projection problem. However, in situations where the original
error operators are associated with quantum operations, the $F_{ij}$'s
do not have a common null space, even after the restrictions of the
previous lemmas have been applied. Otherwise there would be states for
which all $E_i$ have zero probability.

\proof Let $G_1,\ldots, G_k$ be a basis for the linear space spanned
by the $F_{ij}$. We can choose an orthonormal basis of $\cV$
such that in this basis, the matrices $G_i$ have a block form
$[G_{i1},G_{i2},\ldots,G_{ii},\zero,\ldots,\zero]$, where the $G_{ij}$
are $N\times N_j$ matrices of full rank.  We attempt to find the
desired subspace $\cD$ by choosing an orthonormal basis for $\cD$. Let
$X$ be the matrix whose columns are members of this basis.  We wish to
solve the $k$ identities $\alpha_i\one = G_i X$ for $X$ and
$\alpha=(\alpha_i)_i$.  We can write $X$ in block form,
$X=[X_1;\ldots;X_k]$, where $X_i$ is $N_j\times N$ and the $X_i$ are
placed one above the other. The desired identities can be expanded as
\begin{equation}
\label{eq:alphaone}
\alpha_i\one = \sum_{j=1}^i G_{ij}X_j.
\end{equation}
The $X_j$ can be eliminated by solving the equations in order.  That
is, from $\alpha_1\one = G_{11}X_1$ we obtain $\alpha_1=0$ and $X_1=0$
if $N_1\not=N$, and $X_1=\alpha_1G_{11}^{-1}$ otherwise. We write this
as a linear constraint $L_1\cdot\alpha=0$ and an identity
$X_1=\alpha_1\tilde G_{11}$, where $L_1$ may be ``empty'' (if $N_1=N$)
and we set $\tilde G_{11}$ to be any left inverse of $G_{11}$.  Once
we have obtained $X_{j}=\sum_m\alpha_m \tilde G_{jm}$ and linear
constraints $L_j\alpha=0$ for $j<i$, we can solve for $X_i$ by 
substituing in Eq.~\ref{eq:alphaone}:
\begin{equation}
G_{ii}X_i = 
\alpha_i\one-\sum_{j=1}^{i-1}\sum_{m=1}^{j} \alpha_m G_{ij}\tilde G_{jm}.
\end{equation}
The right hand side of this identity is a matrix $H_i$ that depends
linearly on $\alpha$.  $G_{ii}X_i = H_i$ can be solved if and only if
the columns of $H_i$ are in the column span of $G_{ii}$. This
condition yields a set of linear constraints $L_i\alpha = 0$.  If the
constraints are satisfied, then we can compute $X_i = G_{ii}' H_i$,
where $G_{ii}'$ is a left inverse of $G_{ii}$.  We can therefore
define $\tilde G_{im}$ by the identity $X_i=\sum_{m=1}^i
\alpha_j\tilde G_{im}$.  At the end of this process, the only free
variables remaining are the $\alpha_j$, which must be chosen to
satisfy the orthonormality constraint on $X$, $\sum_i X_i^\dagger X_i
= \one$. Expanding, we get
\begin{equation}
\sum_i\sum_{jk}\bar\alpha_j\alpha_k\tilde G_{ij}^\dagger\tilde G_{ik} = 
\one,
\end{equation}
subject to $L_i\alpha = 0$ for all $i$. If the linear
constraints cannot be solved, we are done.
Define $\slb{\rho}{AB}$ by
\begin{equation}
\slb{\rho}{AB} = 
t \sum_i\sum_{jk} \slb{\tilde G_{ij}}{A}{}^\dagger\slb{\tilde G_{ik}}{A}
\ketbras{j}{k}{B},
\end{equation}
where $t$ is chosen so that $\trace(\slb{\rho}{AB})=1$.  Any state
$\kets{\psi}{B}$ in the subspace defined by $L_i\kets{\psi}{B} = 0$
(with $L_i$ defined with respect to the basis consisting of the
$\kets{j}{B}$) that solves the pure $\one$-projection problem yields a
solution for $\alpha$ by letting $\alpha_j$ be a suitably scaled
multiple of the coefficient of $\kets{j}{B}$ in the solution.  It
follows that to complete the proof, it suffices to project
$\slb{\rho}{AB}$ onto the subspace of $\sysfnt{B}$ satisfying the
linear constraints $L_i$ and renormalize the resulting positive
semidefinite operator. This operator is a density matrix for which the
pure $\one$-projection problem is equivalent to the problem of whether
$\sysfnt{I'}$ is protectable.  \qed

The pure $\one$-projection problem may be reduced to a problem
of finding special matrices in a linear space of matrices.

\begin{Theorem}
The pure $\one$-projection problem is polynomially equivalent to
the problem of finding a matrix with orthonormal columns in a linear
space of matrices.
\end{Theorem}

\proof
Consider the pure $\one$-projection problem for $\slb{\rho}{AB}$.  By
purifing $\slb{\rho}{AB}$ with the addition of an environment $E$, we
obtain a pure state $\kets{\psi}{ABE}$ whose reduced density matrix on
$\sysfnt{AB}$ is $\slb{\rho}{AB}$.  The pure $\one$-projection problem
is now equivalent to the problem of finding $\kets{\phi}{B}$ such that
$\bras{\phi}{B}\kets{\psi}{ABE}$ is proportional to a maximally
entangled state between $\sysfnt{A}$ and $\sysfnt{E}$. Note that
without loss of generality, the dimension of $\sysfnt{E}$ is greater
than that of $\sysfnt{A}$. Otherwise, the problem has no solution. We
can expand everything in a basis for the different systems' Hilbert
spaces: $\kets{\phi}{B} = \sum_i \alpha_i\kets{i}{B}$,
$\kets{\psi}{ABE} =
\sum_{ijk}m_{ijk}\kets{i}{A}\kets{j}{B}\kets{k}{E}$.  Let $M_j$ be the
matrix with coefficients $(M_j)_{ki}=m_{ijk}$.  The property that
$\bras{\phi}{B}\kets{\psi}{ABE}$ is maximally entangled is equivalent
to the property that $\sum_j\alpha_j M_j$ has orthonormal columns.

Given any set of matrices $M_j'$ we can reverse the reduction of the
previous paragraph by setting $M_j = t M_j'$ with $t=1/(\sum_j
\trace(M_j'^\dagger M_j'))$ to obtain a state such that its
pure $\one$-projection problem is equivalent to the problem of finding
a matrix with orthonormal columns in the span of the $M_j'$.
\qed

Whether there is an efficient algorithm for finding a matrix with
orthonormal columns in a linear space of matrices is an open question.

\begin{acknowledgments}
Thanks to S.~Glancy and D.~Leibfried for helpful comments and
assistance in preparing this manuscript. This paper is a contribution
of the National Institute of Standards and Technology, an agency of
the U.S. government, and is not subject to U.S. copyright.
\end{acknowledgments}

\raggedright
\bibliography{journalDefs,qc,onns}

\begin{thebibliography}{23}
\expandafter\ifx\csname natexlab\endcsname\relax\def\natexlab#1{#1}\fi
\expandafter\ifx\csname bibnamefont\endcsname\relax
  \def\bibnamefont#1{#1}\fi
\expandafter\ifx\csname bibfnamefont\endcsname\relax
  \def\bibfnamefont#1{#1}\fi
\expandafter\ifx\csname citenamefont\endcsname\relax
  \def\citenamefont#1{#1}\fi
\expandafter\ifx\csname url\endcsname\relax
  \def\url#1{\texttt{#1}}\fi
\expandafter\ifx\csname urlprefix\endcsname\relax\def\urlprefix{URL }\fi
\providecommand{\bibinfo}[2]{#2}
\providecommand{\eprint}[2][]{\url{#2}}

\bibitem[{\citenamefont{Knill et~al.}(2000)\citenamefont{Knill, Laflamme, and
  Viola}}]{knill:qc1999b}
\bibinfo{author}{\bibfnamefont{E.}~\bibnamefont{Knill}},
  \bibinfo{author}{\bibfnamefont{R.}~\bibnamefont{Laflamme}}, \bibnamefont{and}
  \bibinfo{author}{\bibfnamefont{L.}~\bibnamefont{Viola}},
  \bibinfo{journal}{Phys. Rev. Lett.} \textbf{\bibinfo{volume}{84}},
  \bibinfo{pages}{2525} (\bibinfo{year}{2000}).

\bibitem[{\citenamefont{Kribs et~al.}(2005{\natexlab{a}})\citenamefont{Kribs,
  Laflamme, and Poulin}}]{kribs:qc2004a}
\bibinfo{author}{\bibfnamefont{D.}~\bibnamefont{Kribs}},
  \bibinfo{author}{\bibfnamefont{R.}~\bibnamefont{Laflamme}}, \bibnamefont{and}
  \bibinfo{author}{\bibfnamefont{D.}~\bibnamefont{Poulin}},
  \bibinfo{journal}{Phys. Rev. Lett.} \textbf{\bibinfo{volume}{94}},
  \bibinfo{pages}{180501/1} (\bibinfo{year}{2005}{\natexlab{a}}).

\bibitem[{\citenamefont{Viola et~al.}(2001)\citenamefont{Viola, Knill, and
  Laflamme}}]{viola:qc2000c}
\bibinfo{author}{\bibfnamefont{L.}~\bibnamefont{Viola}},
  \bibinfo{author}{\bibfnamefont{E.}~\bibnamefont{Knill}}, \bibnamefont{and}
  \bibinfo{author}{\bibfnamefont{R.}~\bibnamefont{Laflamme}},
  \bibinfo{journal}{J. Phys. A} \textbf{\bibinfo{volume}{34}},
  \bibinfo{pages}{7067} (\bibinfo{year}{2001}).

\bibitem[{\citenamefont{Choi and Kribs}(2005)}]{choi:qc2005b}
\bibinfo{author}{\bibfnamefont{M.-D.} \bibnamefont{Choi}} \bibnamefont{and}
  \bibinfo{author}{\bibfnamefont{D.~W.} \bibnamefont{Kribs}}
  (\bibinfo{year}{2005}), \bibinfo{note}{quant-ph/0507213}.

\bibitem[{\citenamefont{Kribs et~al.}(2005{\natexlab{b}})\citenamefont{Kribs,
  Laflamme, Poulin, and Lesosky}}]{kribs:qc2005a}
\bibinfo{author}{\bibfnamefont{D.~W.} \bibnamefont{Kribs}},
  \bibinfo{author}{\bibfnamefont{R.}~\bibnamefont{Laflamme}},
  \bibinfo{author}{\bibfnamefont{D.}~\bibnamefont{Poulin}}, \bibnamefont{and}
  \bibinfo{author}{\bibfnamefont{M.}~\bibnamefont{Lesosky}}
  (\bibinfo{year}{2005}{\natexlab{b}}), \bibinfo{note}{quant-ph/0504189}.

\bibitem[{\citenamefont{Nielsen and Poulin}(2005)}]{nielsen:qc2005b}
\bibinfo{author}{\bibfnamefont{M.~A.} \bibnamefont{Nielsen}} \bibnamefont{and}
  \bibinfo{author}{\bibfnamefont{D.}~\bibnamefont{Poulin}}
  (\bibinfo{year}{2005}), \bibinfo{note}{quant-ph/0506069}.

\bibitem[{\citenamefont{Zanardi}(1999)}]{zanardi:qc1999c}
\bibinfo{author}{\bibfnamefont{P.}~\bibnamefont{Zanardi}},
  \bibinfo{journal}{Phys. Rev. A} \textbf{\bibinfo{volume}{63}},
  \bibinfo{pages}{012301/1} (\bibinfo{year}{1999}).

\bibitem[{\citenamefont{Alicki}(2004)}]{alicki:qc2004b}
\bibinfo{author}{\bibfnamefont{R.}~\bibnamefont{Alicki}}
  (\bibinfo{year}{2004}), \bibinfo{note}{quant-ph/0411008}.

\bibitem[{\citenamefont{Aliferis et~al.}(2005)\citenamefont{Aliferis,
  Gottesman, and Preskill}}]{aliferis:qc2005b}
\bibinfo{author}{\bibfnamefont{P.}~\bibnamefont{Aliferis}},
  \bibinfo{author}{\bibfnamefont{D.}~\bibnamefont{Gottesman}},
  \bibnamefont{and} \bibinfo{author}{\bibfnamefont{J.}~\bibnamefont{Preskill}}
  (\bibinfo{year}{2005}), \bibinfo{note}{quant-ph/0504218}.

\bibitem[{\citenamefont{Ralph et~al.}(2001)\citenamefont{Ralph, Munro, and
  Milburn}}]{ralph:qc2001b}
\bibinfo{author}{\bibfnamefont{T.~C.} \bibnamefont{Ralph}},
  \bibinfo{author}{\bibfnamefont{W.~J.} \bibnamefont{Munro}}, \bibnamefont{and}
  \bibinfo{author}{\bibfnamefont{G.~J.} \bibnamefont{Milburn}}
  (\bibinfo{year}{2001}), \bibinfo{note}{quant-ph/0110115}.

\bibitem[{\citenamefont{Ralph et~al.}(2003)\citenamefont{Ralph, Gilchrist,
  Milburn, Munro, and Glancy}}]{ralph:qc2003a}
\bibinfo{author}{\bibfnamefont{T.~C.} \bibnamefont{Ralph}},
  \bibinfo{author}{\bibfnamefont{A.}~\bibnamefont{Gilchrist}},
  \bibinfo{author}{\bibfnamefont{G.~J.} \bibnamefont{Milburn}},
  \bibinfo{author}{\bibfnamefont{W.~J.} \bibnamefont{Munro}}, \bibnamefont{and}
  \bibinfo{author}{\bibfnamefont{S.}~\bibnamefont{Glancy}},
  \bibinfo{journal}{Phys. Rev. A} \textbf{\bibinfo{volume}{68}},
  \bibinfo{pages}{042319/1} (\bibinfo{year}{2003}).

\bibitem[{\citenamefont{Shabani and Lidar}(2005)}]{shabani:qc2005a}
\bibinfo{author}{\bibfnamefont{A.}~\bibnamefont{Shabani}} \bibnamefont{and}
  \bibinfo{author}{\bibfnamefont{D.~A.} \bibnamefont{Lidar}}
  (\bibinfo{year}{2005}), \bibinfo{note}{quant-ph/0505051}.

\bibitem[{onn()}]{onns1}
\bibinfo{note}{Strictly speaking these are ``quantum'' subsystems. One can also
  consider classical subsystems. Classical subsystems require an explicit basis
  for encoding classical information. See, for example,~\cite{knill:qc1999b}.}

\bibitem[{\citenamefont{Kempe et~al.}(2001)\citenamefont{Kempe, Bacon, Lidar,
  and Whaley}}]{kempe:qc2001a}
\bibinfo{author}{\bibfnamefont{J.}~\bibnamefont{Kempe}},
  \bibinfo{author}{\bibfnamefont{D.}~\bibnamefont{Bacon}},
  \bibinfo{author}{\bibfnamefont{D.~A.} \bibnamefont{Lidar}}, \bibnamefont{and}
  \bibinfo{author}{\bibfnamefont{K.~B.} \bibnamefont{Whaley}},
  \bibinfo{journal}{Phys. Rev. A} \textbf{\bibinfo{volume}{63}},
  \bibinfo{pages}{042307/1} (\bibinfo{year}{2001}).

\bibitem[{\citenamefont{Poulin}(2005)}]{poulin:qc2005a}
\bibinfo{author}{\bibfnamefont{D.}~\bibnamefont{Poulin}}
  (\bibinfo{year}{2005}), \bibinfo{note}{quant-ph/0508131}.

\bibitem[{\citenamefont{Bacon}(2005)}]{bacon:qc2005b}
\bibinfo{author}{\bibfnamefont{D.}~\bibnamefont{Bacon}} (\bibinfo{year}{2005}),
  \bibinfo{note}{quant-ph/0506023}.

\bibitem[{\citenamefont{Raussendorf and Briegel}(2001)}]{raussendorf:qc2001a}
\bibinfo{author}{\bibfnamefont{R.}~\bibnamefont{Raussendorf}} \bibnamefont{and}
  \bibinfo{author}{\bibfnamefont{H.~J.} \bibnamefont{Briegel}},
  \bibinfo{journal}{Phys. Rev. Lett.} \textbf{\bibinfo{volume}{86}},
  \bibinfo{pages}{5188} (\bibinfo{year}{2001}).

\bibitem[{\citenamefont{Cory et~al.}(1996)\citenamefont{Cory, Fahmy, and
  Havel}}]{cory:qc1996a}
\bibinfo{author}{\bibfnamefont{D.~G.} \bibnamefont{Cory}},
  \bibinfo{author}{\bibfnamefont{A.~F.} \bibnamefont{Fahmy}}, \bibnamefont{and}
  \bibinfo{author}{\bibfnamefont{T.~F.} \bibnamefont{Havel}}, in
  \emph{\bibinfo{booktitle}{Proceedings of the 4th Workshop on Physics and
  Computation}}, edited by \bibinfo{editor}{\bibfnamefont{T.~T.}
  \bibnamefont{{\em et al.}}} (\bibinfo{publisher}{New England Complex Systems
  Institute}, \bibinfo{address}{Boston, Massachusetts}, \bibinfo{year}{1996}),
  pp. \bibinfo{pages}{87--91}.

\bibitem[{\citenamefont{Gershenfeld and Chuang}(1997)}]{chuang:qc1997a}
\bibinfo{author}{\bibfnamefont{N.~A.} \bibnamefont{Gershenfeld}}
  \bibnamefont{and} \bibinfo{author}{\bibfnamefont{I.~L.}
  \bibnamefont{Chuang}}, \bibinfo{journal}{Science}
  \textbf{\bibinfo{volume}{275}}, \bibinfo{pages}{350} (\bibinfo{year}{1997}).

\bibitem[{\citenamefont{Knill and Laflamme}(1998)}]{knill:qc1998c}
\bibinfo{author}{\bibfnamefont{E.}~\bibnamefont{Knill}} \bibnamefont{and}
  \bibinfo{author}{\bibfnamefont{R.}~\bibnamefont{Laflamme}},
  \bibinfo{journal}{Phys. Rev. Lett.} \textbf{\bibinfo{volume}{81}},
  \bibinfo{pages}{5672} (\bibinfo{year}{1998}).

\bibitem[{\citenamefont{Knill and Laflamme}(1997)}]{knill:qc1995e}
\bibinfo{author}{\bibfnamefont{E.}~\bibnamefont{Knill}} \bibnamefont{and}
  \bibinfo{author}{\bibfnamefont{R.}~\bibnamefont{Laflamme}},
  \bibinfo{journal}{Phys. Rev. A} \textbf{\bibinfo{volume}{55}},
  \bibinfo{pages}{900} (\bibinfo{year}{1997}).

\bibitem[{\citenamefont{Hungerford}(1980)}]{hungerford:qc1980a}
\bibinfo{author}{\bibfnamefont{T.~W.} \bibnamefont{Hungerford}},
  \emph{\bibinfo{title}{Algebra}} (\bibinfo{publisher}{Springer Verlag},
  \bibinfo{address}{New York}, \bibinfo{year}{1980}).

\bibitem[{\citenamefont{Struble}(2000)}]{struble:qc2000a}
\bibinfo{author}{\bibfnamefont{C.~A.} \bibnamefont{Struble}}, Ph.D. thesis,
  \bibinfo{school}{Virginia Polytechnic Institute and State University},
  \bibinfo{address}{Blacksburg, Virginia, US} (\bibinfo{year}{2000}),
  \bibinfo{note}{available online at
  \texttt{http://scholar.lib.vt.edu/theses/available/etd-04282000-13520019}}.

\end{thebibliography}

\end{document}